\documentclass[usenatbib]{mnras}
\usepackage{epsfig,graphicx,latexsym}
\usepackage{aas_macros}
\usepackage[fleqn]{amsmath} 
\usepackage{amssymb}  

\newcommand{\be}{\begin{equation}}
\newcommand{\ee}{\end{equation}}
\newcommand{\bel}[1]{\begin{equation}\label{#1}}
\newcommand{\ba}{\begin{eqnarray}}
\newcommand{\ea}{\end{eqnarray}}
\newcommand{\bal}[1]{\begin{eqnarray}\label{#1}}

\newcommand{\pdet}{p_\textrm{det}}
\newcommand{\ppop}{p_\textrm{pop}}

\newcommand{\dd}{\mathrm{d}}
\newcommand{\diff}[2]{\frac{\dd #1}{\dd #2}}

\newcommand{\Ndet}{N_\mathrm{det}}
\newcommand{\Nndet}{N_\mathrm{ndet}}
\newcommand{\Nnobs}{N_\mathrm{nobs}}
\newcommand{\Nobs}{N_\mathrm{obs}}
\newcommand{\Ntotal}{N_\mathrm{total}}
\newcommand{\Kevents}{{\Nobs}}

\newcommand{\vd}{\vec{d}}
\newcommand{\vlambda}{\vec{\lambda}}
\newcommand{\vtheta}{\vec{\theta}}

\usepackage[dvipsnames]{xcolor}

\title[Selection Effects in Hierarchical Modelling]{Extracting distribution parameters from multiple uncertain observations with selection biases}

\author[]{\parbox{\textwidth}{
Ilya Mandel$^{1,2}$\thanks{E-mail: imandel@star.sr.bham.ac.uk},
Will M. Farr$^{3,4}$\thanks{E-mail: will.farr@stonybrook.edu},
Jonathan R. Gair$^5$\thanks{E-mail: J.Gair@ed.ac.uk}
}
\vspace{0.5cm}\\
\parbox{\textwidth}{
$^1$ Monash Centre for Astrophysics, School of Physics and Astronomy, Monash University, Clayton, Victoria 3800, Australia\\
$^2$ Birmingham Institute for Gravitational Wave Astronomy and School of Physics and Astronomy, University of Birmingham,\\
Birmingham, B15 2TT, United Kingdom \\
$^3$ Department of Physics and Astronomy, Stony Brook University, Stony Brook NY 11794, USA \\
$^4$ Center for Computational Astronomy, Flatiron Institute, 162 5th Ave., New York NY 10010, USA \\
$^5$ School of Mathematics, University of Edinburgh, James Clerk Maxwell Building, Peter Guthrie Tait Road, Edinburgh EH9 3FD, UK
}
}

\date{\today}
\pubyear{2018}

\begin{document}

\label{firstpage}
\pagerange{\pageref{firstpage}--\pageref{lastpage}}
\maketitle

\begin{abstract}
We derive a Bayesian framework for incorporating selection effects into population analyses.  We allow for both measurement uncertainty in individual measurements and, crucially, for selection biases on the population of measurements, and show how to extract the parameters of the underlying distribution based on a set of observations sampled from this distribution.  We illustrate the performance of this framework with an example from gravitational-wave astrophysics, demonstrating that the mass ratio distribution of merging compact-object binaries can be extracted from Malmquist-biased observations with substantial measurement uncertainty.
\end{abstract}


\section{Introduction}

The problem of extracting the distributional properties of a population of sources based on a set of observations drawn from that distribution is a common one, frequently labeled as hierarchical modelling \citep[e.g.,][]{Hogg:2010} (\citet{Bovy:2011} call this ``extreme deconvolution'').  In practical applications, one often has to deal with selection effects: the observed population will have a Malmquist bias \citep{Malmquist1922,Malmquist1925} whereby the loudest or brightest sources are most likely to be detected, and it is necessary to correct for this bias in order to extract the true source population \citep[e.g.,][]{ForemanMackey:2014,Farr:2014}.  In other applications, significant measurement uncertainties in the individual observations must be accounted for \citep[e.g.,][]{FarrMandel:2018}.  Of course, these two complications -- measurement uncertainties and selection effects -- are often present simultaneously.

There have been multiple attempts to address the problem of population-based inference with both selection effects and significant measurement uncertainties.  The earliest correct published solution to this problem, as far as we are aware, belongs to \citet{Loredo:2004}.  However, despite the availability of this solution, it is easy to be lured into a seemingly straight-forward but incorrect derivation.   The most common mistake is the modification of the model population distribution to account for the selection function, i.e., the inclusion of the probability of detecting a particular event only as a multiplicative term in the probability of observing that event.   This detection probability is usually included as the probability marginalised over all realisations of the data, ignoring the fact that we know the particular data realisation that has been observed. For a given data realisation the probability that a source is detected, which is a property purely of the data, is by definition equal to one for any data set associated with an observation we are analysing.  On the other hand, as shown below, it is critical to include the detection probability in the normalisation factor to account for the different numbers of events expected to be observed under different population models.

We sketched out the correct approach to including selection effects in \citet{Mandel:2016select} (which is superseded by the present manuscript) and \citet{BBH:O1}.  Other
correct applications in the literature include \citet{2017ApJ...851L..25F}, \citet{Fishbach:2018}, and \citet{Feeney:2019}.  Here, we expand and clarify the earlier treatment of \citet{Loredo:2004} by  presenting two different approaches to solving this problem below: a bottom-up and a top-down derivation, showing that they yield the same result.  Some among us find one or the other approach to be more clear, and we hope that including both will also benefit readers.

We illustrate the derived methodology with two examples.  The first is the classic example of measuring a luminosity function with a flux-limited survey.  The second is an example from gravitational-wave astronomy: the measurement of the mass ratio of merging binary neutron stars.  We show that $\gtrsim 1000$ observations at a signal-to-noise ratio of $\gtrsim 20$ will be necessary to accurately measure the mass ratio distribution.  This feat, which can be accomplished with third-generation ground-based gravitational-wave detectors, could elucidate the details of neutron star formation.

\section{Problem statement and notation}

We consider a population of events or objects, each described by a set of parameters $\vtheta$.  These parameters represent the characteristics of individual events. For example, in the case of compact binary coalescences observed by LIGO and Virgo these would include the masses, spin magnitudes and spin orientations of the two components, the location of the source on the sky, the distance of the source, the orientation and eccentricity of the binary orbit etc. The distribution of events in the population is described via parameters $\vec{\lambda}$, so that the number density of objects follows $\diff{N}{\vtheta} (\vlambda) = N \ppop(\vec{\theta}|\vlambda{}')$. In the gravitational wave-context, these parameters could represent properties of the population like the slope of the mass function of black holes in compact binaries, or the shape of the spin magnitude distribution, or the mixing fractions of different sub-populations. They could also represent physical ingredients used in population synthesis calculations, for example the parameters of the initial mass function, stellar metallicity distribution or stellar winds and the properties of common envelope evolution or of the distribution of supernova kicks. In this second case, the distribution of the individual event properties $\ppop(\vec{\theta}|\vlambda{}')$ could be obtained from the output of population synthesis codes for that particular choice of input physics. We have separated $\vlambda$ into the overall normalisation for the number or rate of events $N$ and the set of parameters describing the shape of the distribution alone $\vlambda{}'$.  For instance, if the underlying distribution is modelled as a multi-dimensional Gaussian, $\vec{\lambda}$ would consist of the mean vector and covariance matrix; alternatively, a non-parametric distribution could be described with a (multi-dimensional) histogram, in which case $\vec{\lambda}$ represents the weights of various histogram bins.

This distribution is sampled by drawing a set of $\Kevents$ ``observed events'' with true parameters $\{\vec{\theta}_i\}$, for $i \in [1,\Kevents]$.   For each object in the population we make a noisy measurement of $\vtheta_i$, represented by a likelihood function relating the measured data, $\vd_i$, to the parameters of the event, $\vtheta$: $p\left( \vd_i \mid \vtheta_i \right)$.

Moreover, based on the observed data, some objects are classed as
``observable'' and others are ``un-observable.''  For example, a survey may
impose a per-pixel or per-aperture threshold on the flux for inclusion of
point-sources in a catalog, or a gravitational wave detector may only report
events whose signal-to-noise ratio rises above some predetermined threshold.
This detection probability can be estimated empirically for a search pipeline via a large injection campaign.
In some cases, it can be modelled analytically; for example, for low-mass compact binaries, the gravitational-wave strain in the frequency domain is proportional to the $5/6$ power of the chirp mass $M_c$, so the detection probability scales as the surveyed volume\footnote{In practice, there are very weak deviations from this power law due to the imperfect -- noisy -- measurement of signal amplitude.}, $\propto M_c^{15/6}$.
Throughout this article, we will assume that whether or not an event is counted as a detection is a property only of the data for each object and so there exists an indicator function $\mathbf{I}(\vec{d})$ that is equal to 1 for ``observable'' objects that would be classified as detections and 0 otherwise; this is by far the most common case for astronomical observations\footnote{An example where the selection may be parameter- rather than data-dependent is in surveys of objects that have been selected based on data in yet other surveys; Maggie Lieu pointed us to X-ray selected populations of galaxy clusters
in a weak-lensing catalog. This can still be treated within the framework proposed here, by considering the combined likelihood for both data sets and marginalising over the ``discarded'' data from the survey used for selection.}.

Our ultimate goal is to determine the population properties $\vec{\lambda}$.  Of course, we cannot uniquely reconstruct $\vec{\lambda}$ using a limited set of observations with selection biases and measurement uncertainties.  The best we can do is compute the posterior probability on $\vec{\lambda}$, the distribution on distributions, given the observations, which, in the usual Bayesian formalism, is given by
\bel{lambda}
p(\vec{\lambda}|\{\vec{d}_i\}) = \frac{p(\{\vec{d}_i\}|\vec{\lambda}) \pi(\vlambda)}{p(\{\vec{d}_i\})},
\ee
where $p(\{\vec{d}_i\}|\vec{\lambda})$ is the likelihood of observing the data set given the population properties, $\pi (\vlambda)$ is the prior on $\vlambda$ and the evidence $p(\{\vec{d}_i\})$ is the integral of the numerator over all $\vlambda$.  This evidence can be used to select between different models for representing the distribution, as in \cite{Farr:2010}.  In the next two sections, we present two alternative ways of deriving $p(\{\vec{d}_i\}|\vec{\lambda})$.

\section{Bottom-up derivation}
\label{sec:inductive}

First, we follow the bottom-up approach of deriving the likelihood for obtaining a particular set of observations given the population parameters, by starting with a simple problem without either measurement uncertainties or selection effects and gradually building up the problem complexity.  For the moment, we assume that we are only interested in the \emph{shape} of the population distribution, and ignore the \emph{normalisation}, or \emph{rate}, of objects in the population; we discuss estimation of both the rate and shape of a population at the end of this section and in \S\, \ref{sec:deductive}.

In the absence of measurement uncertainties, the data can be directly converted into event parameters $\{\vec{\theta}_i\}$, for $i \in [1,\Kevents]$.  The total probability of making this particular set of independent observations is
\bel{obstotal}
p(\{\vec{\theta}_i\}|\vec{\lambda}') =
\prod_{i=1}^\Kevents \frac{\ppop(\vec{\theta}_i|\vec{\lambda}')} {\int \dd\vec{\theta}\, \ppop(\vec{\theta}|\vec{\lambda}')}\, .
\ee
The normalisation factor here accounts for the overall probability of making an observation given a particular choice of $\vec{\lambda}$ (it will be equal to $1$ if $\ppop$ is properly normalised, but we keep the normalisation term for completeness).


In practice, there is often a selection bias involved: some events are easier to observe than others. This can be characterised by a detection probability $\pdet(\vec{\theta})$. We assume for now that when systems are observed their parameters can be measured perfectly, i.e., we directly measure $\{\vec{\theta}_i\}$. This effectively says that the noise in the measurement is negligible\footnote{This is a very artificial model since all detectors have noise and the reason that $\pdet(\vec{\theta})$ is not equal to one is because of that noise. However, it serves to illustrate the basic idea.} and the selection effects can be applied directly to the event parameters: $\pdet(\vec{\theta}) = \mathbf{I}(\vec{\theta})$, i.e., events are either always detected or never detected depending on their parameters.
With the selection effect included,  equation (\ref{obstotal}) becomes \cite[e.g.,][]{Chennamangalam:2012,Farr:2013}
\bal{lambda-selection}
p(\{\vec{\theta}_i\}|\vec{\lambda}')  &=& \prod_{i=1}^\Kevents \frac{ \ppop(\vec{\theta}_i|\vec{\lambda}') \pdet (\vec{\theta}_i)} {\int \dd(\vec{\theta}) \ppop(\vec{\theta}|\vec{\lambda}') \pdet(\vec{\theta})}\nonumber \\ &=& \prod_{i=1}^\Kevents \frac{ \ppop(\vec{\theta}_i|\vec{\lambda}')} {\int \dd(\vec{\theta}) \ppop(\vec{\theta}|\vec{\lambda}') \pdet(\vec{\theta})}\ ,
\ea
where the second equality follows because, by definition, $\mathbf{I}(\vec\theta)=1$ for any event we have included among the set of detections.


In general, we don't have the luxury of directly measuring the parameters of a given event, $\vec{\theta}_i$.  Instead, we measure the data set $\vec{d}_i$ which encodes these parameters but also includes some random noise.  For a given data set and search pipeline, we assume that the detectability is deterministic: if the data exceeds some threshold (e.g., a threshold on the signal to noise ratio), then the event is detectable; otherwise, it's not.  In other words, the detection probability for a given set of parameters introduced earlier is, in fact, an integral over the possible data sets given those parameters:
\be
\pdet (\vec{\theta}) = \int_{\vec{d} > \textrm{threshold}} p(\vec{d}|\vec{\theta}) \dd\vec{d} = \int \mathbf{I}(\vec{d}) p(\vec{d}|\vec\theta) \dd \vec{d}\ .
\label{eq:pdet}
\ee
In the gravitational-wave context, detection is usually well approximated as a cut on the observed signal-to-noise ratio (SNR) and so this detection probability is the likelihood distribution of observed SNRs. There are two stochastic components to the observed SNR --- fluctuations in the detector noise which change the observed SNR relative to the intrinsic SNR, and fluctuations in the intrinsic SNR due to variations in the source parameters.  For an example of the latter, the expected signal amplitude is a strong function of the mass --- a selection effect that is critical to consider when inferring the underlying distribution of binary black hole masses from the observed events~\cite{BBH:O1,2017ApJ...851L..25F}. As another example, the intrinsic SNR also depends on extrinsic parameters of the binary, i.e., the sky location and orientation of the system. That dependence is largely encoded in the distribution of the parameter $\Theta$ described in~\cite{FinnChernoff:1993}. The function $\pdet (\vec{\theta})$ encodes both these types of intrinsic selection effect, plus marginalisation over instrumental noise fluctuations.

Using Eq.~(\ref{eq:pdet}), we can write the probability of observing a particular data set (where ``observing'' implies that the data are above the threshold, hence included as one of our $k$ observations) given the assumed distribution parametrised by $\vec{\lambda}'$ as
\bel{pdlambda}
p(\vec{d}|\vec{\lambda}') = \frac{\int \dd\vec{\theta} p(\vec{d}|\vec{\theta}) \ppop(\vec{\theta}|\vec{\lambda}')}{\alpha(\vec{\lambda}')}\, ,
\ee
where the normalisation factor $\alpha(\vec{\lambda}')$ is given by
\ba
\alpha(\vec{\lambda}') &\equiv& \int_{\vec{d} > \textrm{threshold}} \dd\vec{d} \, \int \dd\vec{\theta} p(\vec{d}|\vec{\theta}) \ppop(\vec{\theta}|\vec{\lambda}')\nonumber\\
&=&  \int \dd\vec{\theta} \left[\int_{\vec{d} > \textrm{threshold}} \dd\vec{d} p(\vec{d}|\vec{\theta})\right]  \ppop(\vec{\theta}|\vec{\lambda}')\nonumber\\
&\equiv& \int \dd\vec{\theta} \pdet(\vec{\theta}) \ppop(\vec{\theta}|\vec{\lambda}')\, .
\ea
This normalisation factor can be interpreted as the fraction of events in the Universe that would be detected for a particular population model, characterised by the population parameters $\vec\lambda'$.

Thus, in the presence of both measurement uncertainty and selection effects, equations (\ref{obstotal}) and (\ref{lambda-selection}) become:
\bel{lambdauncertain}
p(\{\vec{d}_i\}|\vec{\lambda}')  = \prod_{i=1}^\Kevents \frac
{\int \dd\vec{\theta} p(\vec{d}_i|\vec{\theta}) \ppop(\vec{\theta}|\vec{\lambda}')}
{\int \dd\vec{\theta} \pdet(\vtheta) \ppop(\vec{\theta}|\vec{\lambda}')} \, .
\ee
The presence of the likelihood $ p(\vec{d}_i|\vec{\theta})$ in this equation is a reminder that we do not have a perfect measurement of the parameters of a given event.  The likelihood can be rewritten in terms of the posterior probability density function (PDF) $p(\vec{\theta}_i|\vec{d}_i)$ that is computed in the course of single-event parameter estimation using some assumed prior $\pi(\vec{\theta})$:
\be
p(\vec{d}_i|\vec{\theta}_i) = \frac{p(\vec{\theta}_i|\vec{d}_i) p(\vec{d}_i)} {\pi(\vec{\theta})}\, .
\ee
Thus, each term of the product in Eq.~(\ref{lambdauncertain}) is a normalised convolution integral of the population with the posterior PDF \citep{Mandel:2010stat}.

In practice, the posterior PDF $p(\vec{\theta}_i|\vec{d}_i)$ is often discretely sampled with $S_i$ samples from the posterior, $\{^j\vec{\theta}_i\}$, for $j \in [1,S_i]$.   Because the samples are drawn according to the posterior, the parameter space volume associated with each sample is inversely proportional to the local PDF,  $d^j\vec{\theta}_i \propto \left[p(^j\vec{\theta}_i | \vec{d}_i)\right]^{-1}$.
This allows us to easily replace the integral in Eq.~(\ref{lambdauncertain}) with a discrete sum over PDF samples:
\bel{lambda-discrete}
p(\{\vec{d}_i\}|\vec{\lambda}') = \prod_{i=1}^\Kevents  \frac
{\frac{1}{S_i} \sum_{j=1}^{S_i} \ppop(^j\vec{\theta}_i|\vec{\lambda}') \frac{p(\vec{d}_i)}{\pi(\vec{\theta})}}
{\int \dd\vec{\theta} \pdet(\vec{\theta}) \ppop(\vec{\theta}|\vec{\lambda}')} \ .
\ee

Finally, the posterior on the underlying population parameters $\vec{\lambda}'$ is given by substituting equation (\ref{lambda-discrete}) into equation (\ref{lambda}):
\bal{lambda-total}
p(\vec{\lambda}'|\{\vec{d}_i\}) &=& \frac{\pi(\vec{\lambda}')}{p(\{\vec{d}_i\})}
\prod_{i=1}^\Kevents\frac
{\frac{1}{S_i} \sum_{j=1}^{S_i} \frac{\ppop(^j\vec{\theta}_i|\vec{\lambda}')}{\pi(\vec{\theta})}p(\vec{d}_i)}
{\int \dd\vec{\theta} \pdet(\vtheta) \ppop(\vec{\theta}|\vec{\lambda}')} \nonumber\\
&=&
\pi(\vec{\lambda}') \prod_{i=1}^\Kevents\frac
{\frac{1}{S_i} \sum_{j=1}^{S_i} \frac{\ppop(^j\vec{\theta}_i|\vec{\lambda}')}{\pi(\vec{\theta})}}
{\int \dd\vec{\theta} \pdet(\vtheta) \ppop(\vec{\theta}|\vec{\lambda}')} \ .
\ea

Of course, if interested in the distribution of a single parameter, we can marginalise over Eq.~(\ref{lambda-total}) in the usual way, by integrating over the remaining parameters.

We have so far described inference based on the shape of the distribution $\ppop(\vec{\theta}|\vec{\lambda}')$ while ignoring the overall normalisation.  This is appropriate when the overall normalisation on the population counts is not interesting, or when the bulk of the information comes from the distribution properties rather than the detection rate (a single data point). This is a reasonable assumption in the gravitational-wave context, where the astrophysical uncertainty on the rates of compact object mergers covers several order of magnitude. While inferring the rate is of great interest, models may not predict it with sufficient precision for that measurement to have strong constraining power.

In contexts in which the expected number of detections $\Ndet$ can be predicted, this can be readily included in the framework. The probability of observing $k$ events is given by the Poisson distribution
\be
p(k|\Ndet) = e^{-\Ndet} (\Ndet)^\Kevents\ .
\ee
Here, the usual $\Kevents!$ term in the denominator is absent because the events are distinguishable by their data; in any case, as a normalisation term that depends on the data only, it would not impact inference on $\vec{\lambda}$.  The expected number of detections once selection effects are included is (cf.~Eq.~(\ref{eq:Ndet})):
\bel{eq:Ndet3}
\Ndet \left( \vlambda \right) \equiv \int_{ \vd > \textnormal{threshold} } \dd \vd \, \dd \vtheta \, p( \vd | \vtheta) \frac{\dd N}{\dd \vtheta}(\vlambda) = N \alpha(\vlambda{}') \ .
\ee
The posterior on the population parameters with the rate included becomes
\bal{lambda-total-rate}
p(\vlambda{}', N|\{\vec{d}_i\}) =
\pi(\vlambda{}') \pi(N) \prod_{i=1}^\Kevents\frac
{\frac{1}{S_i} \sum_{j=1}^{S_i} \frac{\ppop(^j\vec{\theta}_i|\vlambda{}')}{\pi(\vec{\theta})}}
{\int \dd\vec{\theta} \pdet(\vtheta) \ppop(\vec{\theta}|\vlambda{}')} \nonumber \\ \times e^{-\Ndet} (\Ndet)^\Kevents \ .
\ea
Note that if a prior $\pi(N) \propto 1/N$ is assumed on the intrinsic event number or rate \citep{Fishbach:2018}, equation (\ref{lambda-total-rate}) can be marginalised over $N$ to again yield Eq.~(\ref{lambda-total}) up to a normalisation constant, which depends only on the number of observed events and would not impact inference on model parameters:
\bal{rate-marg}
\int \dd N  \frac{\pi(\vlambda{}')}{N} \prod_{i=1}^\Kevents\frac {\sum_{j=1}^{S_i} \frac{\ppop(^j\vec{\theta}_i|\vlambda{}')}{\pi(\vec{\theta})}}{S_i\ \alpha(\vlambda{}')} e^{-N \alpha(\vlambda{}')} \left(N \alpha(\vlambda{}')\right)^\Kevents \nonumber \\ = (\Kevents-1)! \
\pi(\vlambda{}') \prod_{i=1}^\Kevents\frac {\sum_{j=1}^{S_i} \frac{\ppop(^j\vec{\theta}_i|\vlambda{}')}{\pi(\vec{\theta})}}{S_i\ \alpha(\vlambda{}')}
\ea
where we used
\bal{norm}
\int \frac{\dd N}{N} e^{-N \alpha(\vlambda{}')} \left(N \alpha(\vlambda{}')\right)^\Kevents &=& \int \dd\Ndet e^{-\Ndet} \Ndet^{\Kevents-1} \nonumber\\
&=&  \Gamma(k) = (\Kevents-1)!
\ea

\section{Top-down derivation}
\label{sec:deductive}

Alternatively, we consider a top-down calculation. If we have observed a
representative sample from the population (i.e.\ a ``fair draw''), then the
appropriate (unnormalised) joint distribution for the parameters $\left\{
\vtheta_i \right\}_{i=1}^{\Ntotal}$ and observations $\left\{ \vd_i \right\}$ of
the $i = 1, \ldots, \Ntotal$ objects given the parameters $\vlambda$ describing
the population (again, $\vlambda$ are \emph{all} parameters describing the
population, including the rate, while $\vlambda{}'$ are parameters that only
describe the \emph{shape} of the population) is
\begin{equation}
  \pi\left(\left\{ \vtheta_i \right\}, \left\{ d_i \right\} \mid \vlambda \right) \propto \left[ \prod_{i=1}^{\Ntotal} p\left( \vd_i \mid \vtheta_i \right) \diff{N}{\vtheta_i}\left( \vlambda \right) \right] \exp\left[ - N\left( \vlambda \right) \right],
\end{equation}
where
\begin{equation}
N\left( \vlambda \right) \equiv \int \dd \vd \, \dd \vtheta p\left( \vd \mid \vtheta \right) \diff{N}{\vtheta}\left( \lambda \right)
\end{equation}
is the expected number of objects in the population\footnote{The rationale for
writing this as a double-integral, when the integral over $\vd$ is in fact
trivial -- since the likelihood is normalised over $\vd$ -- will become apparent
below.}.  This is the standard likelihood for a hierarchical analysis of an
inhomogeneous Poisson process
\citep{Loredo:1995,Hogg:2010,Mandel:2010stat,Youdin:2011,ForemanMackey:2014,Farr:2013,Barrett:2017FIM}.

If some objects are classed as  ``observable'' (indexed by $i$) and others are
``un-observable'' (indexed by $j$), the complete set of observations partitions
into two subsets of cardinality $\Nobs$ and $\Nnobs$:
\ba
\label{eq:selected-unselected-partition-posterior}
  \pi\left(\left\{ \vtheta_i \right\}, \left\{ \vtheta_j \right\}, \left\{ d_i \right\}, \left\{ d_j \right\} \mid \vlambda \right)  \propto \left[ \prod_{i=1}^{\Nobs} p\left( \vd_i \mid \vtheta_i \right) \diff{N}{\vtheta_i}\left( \vlambda \right) \right] \nonumber \\ \times \left[ \prod_{j=1}^{\Nnobs} p\left( \vd_j \mid \vtheta_j \right) \diff{N}{\vtheta_j}\left( \vlambda \right) \right] \exp\left[ - N\left( \vlambda \right) \right].
\ea
Again, a key point is that we can perform this partitioning simply by examining
the \emph{data} obtained for each object.

It is common for the data associated with ``non-observable'' objects to be
completely \emph{censored}; that is, it often does not appear in a catalog or
otherwise at all.  In this case, it is appropriate to marginalise over the
parameters and (unknown) data for the ``non-observable'' objects.  Doing so
destroys the distinguishability inherent in the inhomogeneous Poisson
distribution, so we must introduce a factor of $\Nnobs!$ to account for the
over-counting:
\ba
  \pi \left(\left\{ \vtheta_i \right\}, \left\{ d_i \right\},\Nnobs \mid \vlambda \right)  \propto  \left[ \prod_{i=1}^{\Nobs} p\left( \vd_i \mid \vtheta_i \right) \diff{N}{\vtheta_i}\left( \vlambda \right) \right] \nonumber \\  \times \frac{\Nndet^{\Nnobs}\left( \vlambda \right)}{\Nnobs!} \exp\left[ - N\left( \vlambda \right) \right],
\ea
where
\begin{equation}
\Nndet\left( \vlambda \right) \equiv \int_{\left\{ \vd \mid \textnormal{non-detection} \right\}} \dd \vd \, \dd \vtheta \, p\left( \vd \mid \vtheta \right) \diff{N}{\vtheta}\left( \vlambda \right)
\end{equation}
is the expected number of non-detections in the population model.  Stopping here
we would have a model similar to the ones discussed in \citet{Messenger2013}
(though that reference did not discuss rate
estimation); however, it is  common to not even know \emph{how many}
non-detected objects there were in a given survey or data set.  In this case we
must marginalise -- sum, since counting is a discrete operation -- over the
unknown number of non-detections, $\Nnobs$, yielding
\ba
\pi\left(\left\{ \vtheta_i \right\}, \left\{ d_i \right\} \mid \vlambda \right) & \propto & \left[ \prod_{i=1}^{\Nobs} p\left( \vd_i \mid \vtheta_i \right) \diff{N}{\vtheta_i}\left( \vlambda \right) \right]  \\ & \times & \exp\left[ - \left( N\left( \vlambda \right) - \Nndet\left( \vlambda \right) \right) \right], \nonumber
\ea
or
\begin{equation}\label{lambda-total-rate4}
  \pi\left(\left\{ \vtheta_i \right\}, \left\{ d_i \right\} \mid \vlambda \right) \propto \left[ \prod_{i=1}^{\Nobs} p\left( \vd_i \mid \vtheta_i \right) \diff{N}{\vtheta_i}\left( \vlambda \right) \right] \exp\left[ - \Ndet\left( \vlambda \right) \right],
\end{equation}
where $\Ndet$ -- the compliment of $\Nndet$ -- is the expected number of
detections under the population model:
\bal{eq:Ndet}
  \Ndet\left( \vlambda \right) &\equiv& \int_{\left\{ \vd \mid \textnormal{detection} \right\}} \dd \vd \, \dd \vtheta \, p\left( \vd \mid \vtheta \right) \diff{N}{\vtheta}\left( \vlambda \right)\nonumber \\
  &=& \int  \dd \vd \, \dd \vtheta \, \mathbf{I}(\vec{d})  p\left( \vd \mid \vtheta \right) \diff{N}{\vtheta}\left( \vlambda \right)\ .
\ea
This equation is the posterior for a hierarchical analysis of the number density
and properties of objects from a data set subject to selection effects
\citep[e.g.][]{Gair:2010,Youdin:2011,Fishbach:2018,Wysocki:2018}.

This is the same result we derived in \S\, \ref{sec:inductive}. Each multiplicative term in the numerator of Eq.~(\ref{lambda-total-rate}) from  \S\, \ref{sec:inductive} is the integral $\int \dd\vec{\theta} p(\vec{d}_i|\vec{\theta}) \ppop(\vec{\theta}|\vec{\lambda}')$, approximated as a Monte Carlo sum over the posterior samples.  The denominator of Eq.~(\ref{lambda-total-rate}) is $\alpha^{\Nobs}$.  Meanwhile,  $\alpha = \Ndet/N$ according to equation (\ref{eq:Ndet3}), which is identical to equation (\ref{eq:Ndet}) from this section.  With the substitution $\ppop(\vec{\theta}|\vec{\lambda}') = (dN/d\vtheta) / N$, the entire fraction in Eq.~(\ref{lambda-total-rate}) is identical to the first term of Eq.~(\ref{lambda-total-rate4}) divided by $\Ndet^{\Nobs}$, which cancels the last term of Eq.~(\ref{lambda-total-rate}).  Thus, we see that equations (\ref{lambda-total-rate}) and (\ref{lambda-total-rate4}) are equivalent up to the choice of priors.

As in \S\, \ref{sec:inductive}, if we re-parameterise $\diff{N}{\vtheta}$ so that we can write
\begin{equation}
  \diff{N}{\vtheta} \equiv N p\left( \vtheta \mid \vlambda{}' \right)
\end{equation}
with $p\left( \vtheta \mid \vlambda{}'\right)$ integrating to 1 over the
population for any value of the new parameters $\vlambda{}'$, impose a prior
$p\left( N \right)\propto 1/N$, and marginalise over $N$, we arrive at the
treatment of selection functions for estimating population distributions from
\citet{Loredo:2004,BBH:O1}.
This correspondence only holds with a $1/N$ scale-invariant prior on the number of objects in the population (see \citet{Fishbach:2018} and Eq.~(\ref{rate-marg}) above); other priors are,
of course, possible, but will not marginalise to the population analysis above.

Note that the commonly-employed technique of modifying $\diff{N}{\vtheta}$ to
account for the selection function is not correct, and will lead to biased
results as long as the selection is dependent only on the observed
data.

\section{How important is it to include selection effects?}\label{sec:numevs}
It is natural to ask how many events you will need to observe before the incorrect treatment of selection effects starts to influence the results. Any incorrect analysis, i.e., writing down a posterior distribution that is not consistent with the true data generating process, will lead to a bias in the result and might also change the inferred posterior uncertainty. Asymptotically, the bias remains constant while the uncertainty decreases like the square root of the number of events. Therefore after sufficient observations have been made the result will be inconsistent with the true parameter values. The number of events that can be observed before the bias becomes important depends both on what particular ``wrong method'' is being used and on the specific problem under consideration. One plausible wrong method is that selection effects will be ignored completely, but more often selection effects are included in an incorrect way. For example, one might write down the likelihood for an individual detected event as
$$
\int p(\vec{d}, {\rm det} | \vec\theta) p_{\rm pop}(\vec\theta | \vec\lambda') {\rm d}\vec\theta
$$
which acknowledges that we have only used detected events (indicated by the flag ``det''). Then an incorrect assumption is made that the specific data generation process and the question of whether or not the event is detected are independent, so that the first term can be factorised as
$$
\int p(\vec{d} | \vec\theta) p_{\rm det} (\vec\theta) p_{\rm pop}(\vec\theta | \vec\lambda'){\rm d}\vec\theta.
$$
This differs from the true result in two ways --- the normalisation term $1/\alpha(\vec\lambda)$ is missing, and there is an extra factor of $p_{\rm det}(\vec\theta)$ in the numerator.

A slightly more astute practitioner might realise that the selection bias modifies the probability distribution for the parameters of observed events so that this becomes
$$
p(\vec\theta | {\rm det}, \vec\lambda') = \frac{p_{\rm det}(\vec\theta) p_{\rm pop}(\vec\theta|\vec\lambda')}{\alpha(\vec\lambda')}
$$
but then fail to also condition the likelihood $p(d|\vec\theta)$ on detection and use
$$
\frac{1}{\alpha(\vec\lambda')} \int p(d|\vec\theta){p_{\rm det}(\vec\theta) p_{\rm pop}(\vec\theta|\vec\lambda')} {\rm d} \vec\theta
$$
which includes the correct normalisation factor but still has the additional $p_{\rm det}(\vec\theta)$ in the numerator. In this latter case, the differences will only become apparent once a sufficient number of events with $p_{\rm det}(\vec\theta)$ significantly different from $1$ have been observed. The number of events required would scale like the inverse of the fraction of the observable parameter space where selection effects are important, although the exact number of events would also depend on how much information those events contained about $\vec\lambda'$, i.e., how much the properties of those events depend on the properties of the population.

In the former case, every event contributes to a mistake in inference as the factor $1/\alpha(\vec\lambda')$ is also missing. The number of events required before the error becomes apparent will then depend on how strongly this varies with the population parameters, which depends on the particular inference problem. For example, in the case of inferring the slope of the black hole mass function from binary black hole mergers observed by LIGO, this would be a strong effect as shallower mass functions give more higher-mass events, which are visible to greater distances and so a higher proportion of the total population lies within the LIGO detector horizon (see, for example,~\cite{2017ApJ...851L..25F}). However, in the case of inferring the Hubble constant using binary neutron star observers with counterparts, the natural prior on the distance distribution is uniform in comoving volume and, since mass redshifting and non-Euclidean cosmological corrections are negligible within the current LIGO horizon, the selection effect is largely independent of the Hubble constant~\cite{GW170817H0Paper}. To be concrete, in the example that will be described in the next section, we repeated the analysis using the former of these wrong methods (as a worst-case scenario) and we show the results of that analysis as dashed lines in Figure~\ref{fig:ppplot}. That figure shows the probability-probability plot, i.e., the fraction of times the true parameters lie at a particular significance level over many experiments. For true and modelled distributions that are both Gaussians with common variance $\sigma$ but means that differ by a bias $b$, the amount by which the p-p plot deviates from the diagonal depends on $b/\sigma$ (see discussion in~\cite{JGCMppplot}). We see that, for that specific example, with $10$ events the bias is already evident in the p-p plot, but at a level consistent with $b/\sigma < 1$. So, there is a bias but it is smaller than the typical statistical error. For 100 events the effect is much more pronounced and consistent with $b/\sigma \sim$a few, so for $100$ events the result will be appreciably biased. These numbers are for a specific problem and the threshold for inclusion of selection effects to avoid bias will vary from problem to problem. It is therefore important to always include selection effects properly in the analysis, unless there is a good reason to believe that they can be ignored, which typically could only be assessed by doing the analysis including selection effects anyway.

\section{An illustration: measuring a luminosity function with a flux-limited survey}
\label{sec:luminosity}

Measuring a luminosity function from a flux-limited survey is a
classic problem in astronomy that deals with selection effects (see, e.g.,
\cite{Malmquist1922}).  Here we apply the method discussed in the previous
sections to a toy-model, but illustrative, version of this problem.

Suppose the luminosity function of our objects can be modeled by a Schechter
function \citep{Schechter}:
\begin{equation}
  \label{eq:luminosity-function}
  \diff{N}{L} = \frac{\Lambda}{L_*^{1+\alpha} \Gamma\left(1+\alpha\right)} L^\alpha \exp\left[ - \frac{L}{L_*}\right]
\end{equation}
with $\alpha > -1$ and $L_*>0$ parameters controlling the shape of the
distribution and $\Lambda$ the expected number of objects in the survey volume
(i.e. the overall normalization).

\begin{figure}\center
  \includegraphics[width=\columnwidth]{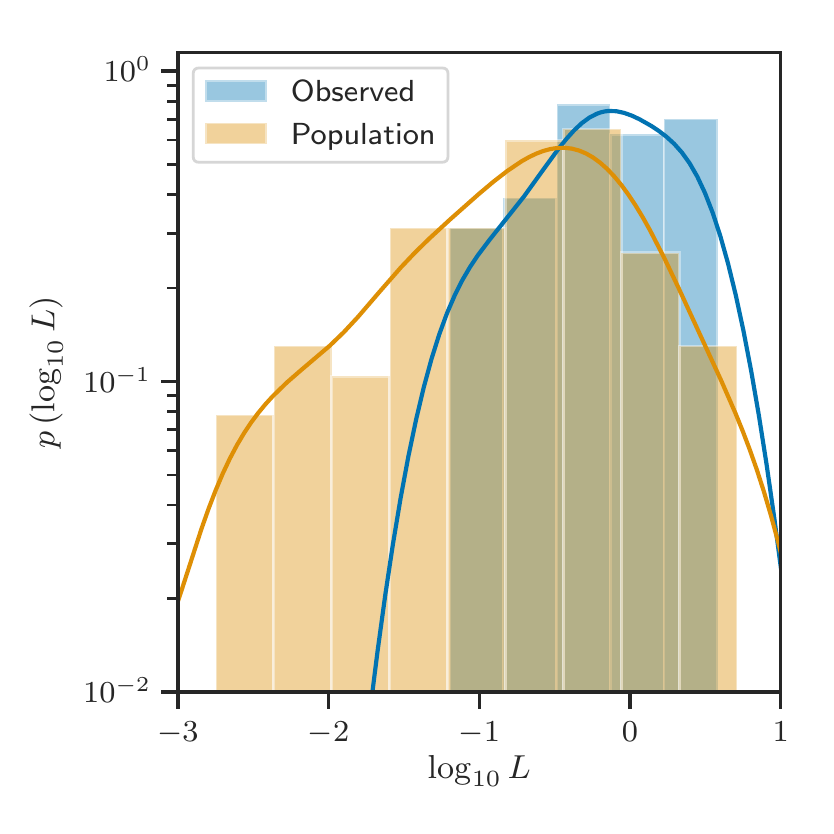}
  \caption{The distribution of observed (blue) and true (orange) luminosities for
  a draw from the model discussed in Section \ref{sec:luminosity}.  Due to
  selection effects, the distribution of observed luminosities peaks at higher
  luminosity and falls more rapidly at low luminosity than the true distribution
  of sources.}
\label{fig:luminosities}
\end{figure}

Somewhat unrealistically, we suppose we can measure distances to objects
perfectly, but that we typically measure fluxes (and therefore luminosities)
with $\sigma_L \simeq 5\%$ uncertainty and that the measurement process results
in a log-normal likelihood function:
\begin{multline}
p\left( L_\mathrm{obs} \mid L \right) = \\ \frac{1}{\sigma_L L_\mathrm{obs} \sqrt{2\pi}} \exp\left[ - \frac{1}{2} \left(\frac{\log L - \log L_\mathrm{obs}}{\sigma_L} \right)^2\right].
\end{multline}
We assume a Euclidean universe, so in appropriate units a flux limit for
detection of $F_\mathrm{th}$ implies a probability of detection for an observed
luminosity of
\begin{equation}
P_\mathrm{det} \left( L_\mathrm{obs} \right) = \begin{cases}
1 & \frac{L_\mathrm{obs}}{4 \pi z^2} > F_\mathrm{th} \\
0 & \mathrm{otherwise}
\end{cases},
\end{equation}
where $z$ is the redshift (distance) to the object.  For computational
efficiency, we assume that our objects are uniformly distributed in $z$ for $0
\leq z \leq 2$ (this assumption reduces the number of un-observable objects
compared to a more realistic volumetric distribution).  We choose true values of
the parameters in this model to be $\Lambda = 100$, $L_* = 1$, $\alpha = -1/2$,
and $F_\mathrm{th} = 1/4\pi$; this latter choice means that the detection
probability for a $L_*$ object at $z = 1$ is 50\%.  For these choices, one draw
of a random universe produces the distribution of observed and true luminosities
shown in Figure \ref{fig:luminosities}.  In this particular draw, we observed 24
objects and missed 80 in our survey.

\begin{figure}\center
  \includegraphics[width=\columnwidth]{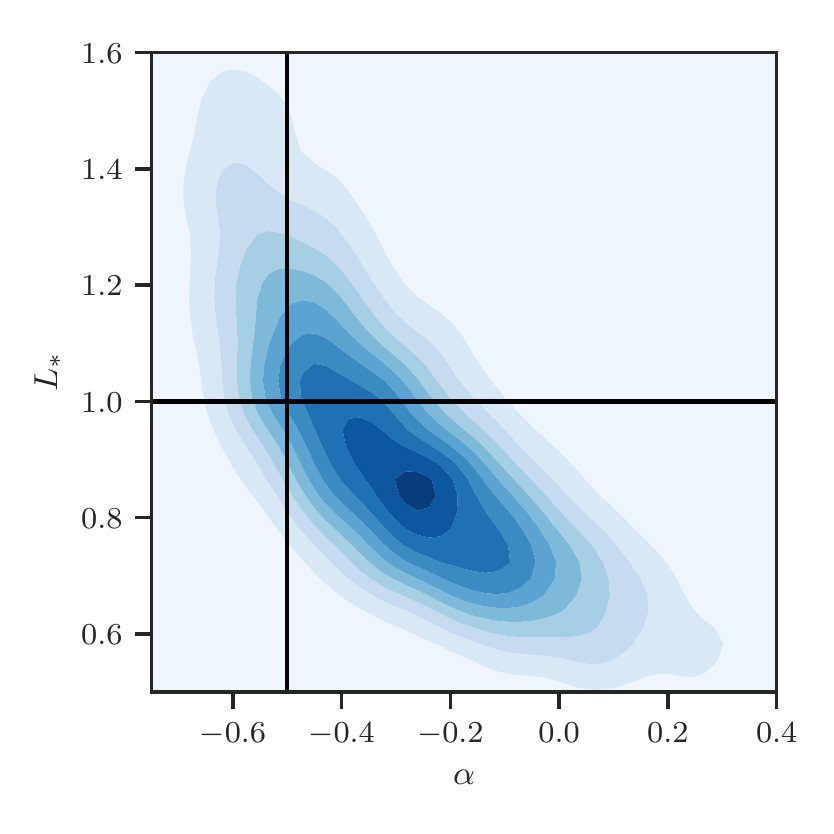}
  \caption{Marginal posterior distribution for the $L_*$ and $\alpha$ parameters
  of the luminosity function (see Eq.\ \eqref{eq:luminosity-function}) from the
  model and data described in Section \ref{sec:luminosity}.  Black lines
  indicate the true values of the parameters.}
\label{fig:Lstar-alpha}
\end{figure}

Applying the ``top-down'' methodology to this problem, the crucial integral in
Eq.\ \eqref{eq:Ndet} is not analytically tractable, though both the population
distribution and the selection function are simple functions.  We must evaluate
this integral numerically.  We choose to do this by sampling over the
un-observed population and associated data (subject to the constraint that the
fluxes associated to the un-observed population are always below
$F_\mathrm{th}$) in a MCMC at the same time as we sample the properties of the
population and observed objects.  That is, we explicitly implement Eq.\
\eqref{eq:selected-unselected-partition-posterior} as our posterior density,
summing over the (unknown) number of non-detected systems.  Sampling over the
un-observed population with this posterior is a method for numerically
evaluating the selection integral.  Code and results implementing this model in
the \texttt{stan} sampler \citep{STAN} can be found at
\url{https://github.com/farr/SelectionExample}.  One result of the sampling is
an estimate of the luminosity function parameters $L_*$ and $\alpha$; a joint
posterior on these parameters appears in Figure \ref{fig:Lstar-alpha}.  The
analysis also recovers with similar accuracy the expected number of objects in
the survey volume ($\Lambda$), improved estimates of each object's intrinsic
luminosity (informed by the population model), and luminosity distributions
of the set of objects too dim to be observed by the survey, as a by-product of
the selection function modelling.

\section{An illustration: measuring the mass ratio of binary neutron stars}

Do all or the majority of merging binary neutron stars have mass ratios very close to unity? Is the answer to this question redshift- or metallicity-dependent?  This question is an important science driver for third-generation gravitational-wave detectors\footnote{This was identified as a key goal during an ongoing study commissioned by the Gravitational Wave International Committee (GWIC), \url{https://gwic.ligo.org/3Gsubcomm/charge.shtml}.}.   Here, we examine how many neutron star binary mergers must be detected in order to measure the mass-ratio distribution, providing an illustration of the methodology described in the previous sections.

The binary neutron star mass ratio distribution is sensitive to the mass ejections associated with neutron star formation in a supernova and the velocity kicks that neutron stars receive at birth.   For example, figure 3 of \citet{VignaGomez:2018} illustrates the differences in the mass ratio distributions under different assumptions about mass fallback and natal kicks.  Since models show a preference for equal mass ratios $q=m_2/m_1$, we assume a simple single-parameter form for the intrinsic mass ratio distribution:
\bel{qform}
p(\eta) \propto e^{(\eta-0.25)/\lambda}\, ,
\ee
where $\eta=q/(1+q)^2$ is the symmetric mass ratio.  We use the symmetric mass ratio because it tends to have more symmetric error bars than $q$; when component masses are equal, $q=1$ and $\eta=0.25$.

The likelihood function on the data is, in general, quite complex \citep{Veitch:2014}, and depends on a multitude of other parameters, such as spins, which must then be marginalised over to obtain $p(\vec{d}|\eta)$.  We will approximate the problem by viewing the data as a point estimate of the symmetric mass ratio $\hat{\eta}$ (one can think of it as a maximum-likelihood estimate) with a Gaussian likelihood function given by
\be
p(\hat{\eta}|\eta) \propto \exp \left\{{-\frac{(\hat{\eta}-\eta)^2}{2 \sigma_\eta^2}}\right\}.
\ee
We use a simple Fisher-information-matrix analysis with a noise power spectral density shape representative of a potential third-generation detector\footnote{We assume that the noise spectral density is proportional to the LIGO A+ design, \url{https://dcc.ligo.org/LIGO-T1800042/public}.} to estimate the expected measurement uncertainty $\sigma_\eta$.   We follow \citet{PoissonWill:1995} in using frequency-domain post-Newtonian waveforms, which can be analytically differentiated and are adequate for binary neutron star analysis, allowing us to rapidly estimate the accuracy of inference.   We do not impose priors, include a spin-orbit coupling term but ignore the spin-spin coupling term as suggested by \citet{PoissonWill:1995}.  We  derive the following simple fit to the measurement uncertainty on $\eta$ for a signal from a canonical $1.4+1.4\ M_\odot$ binary with non-spinning components as a function of the event signal-to-noise ratio $\rho$:
\bel{sigmaeta}
\sigma_\eta = \frac{0.12}{\rho}+\frac{4}{\rho^2}+\frac{250}{\rho^3}
\ee
This fit is accurate to better than 10\% for $\rho>18$.  The inverse of the Fisher information matrix is no longer a good estimate for the covariance matrix at lower values of $\rho$ where the linear signal approximation breaks down, the log-likelihood ceases to be well approximated by a quadratic \citep{Vallisneri:2008}, and the prior constraints on variables strongly correlated with $\eta$, such as the spin parameters, become increasingly important.  In any case, the mass ratio constraints become very poor at low $\rho$; for example, despite a $\rho$ of 32, the mass ratio of the binary neutron star merger GW170817 could only be constrained to $q \in [0.4, 1]$ at 90\% confidence \citep{GW170817}.

The signal-to-noise ratio at a given distance scales as $M_c^{5/6}$, where $M_c \propto \eta^{3/5}$ is the chirp mass, consistent with the inspiral amplitude scaling.  We assume that the distance $D$ to the event is drawn from a $p(D) \propto D^2$ distribution consistent with a flat, isotropic universe, and is known perfectly.  With this simplification, the signal-to-noise ratio as used in Eq.~(\ref{sigmaeta}) follows
\be
 \rho \propto \eta^{0.5}  \frac{18}{D}\, .
\ee
The observed signal-to-noise ratio $\hat{\rho}$ follows the same scaling, but with the dependence on the data $\hat{\eta}$, not the true event mass ratio $\eta$.  In line with comments on the validity of the Fisher information matrix we will use a detection threshold $\hat{\rho} \geq 18$ in this simplified treatment; the detectability conditioned on the observed data is thus independent of the source properties.

\begin{figure}\center
\hspace{-0.1in}
  \includegraphics[width=1.1\columnwidth]{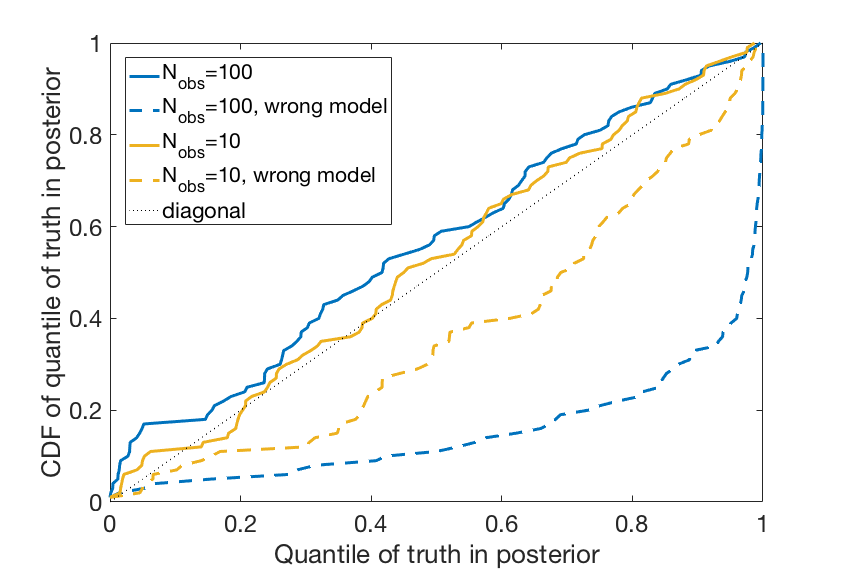}
  \caption{The p-p plot of the cumulative distribution of the quantile of the true value of $\lambda$ within its posterior as estimated from 10 (solid orange curve) and 100 (solid blue curve) mock data sets. These are consistent with the diagonal (dashed black line). For comparison we show the corresponding results, as dashed lines, from using one particular wrong method, as described in Section~\ref{sec:numevs}.}
\label{fig:ppplot}
\end{figure}

We test the self-consistency of the inference on $\lambda$ by creating 100 mock populations with random values of $\lambda$ drawn from the flat prior $\lambda \in [0, 0.1]$.  For each population, we compute the posterior distribution on $\lambda$ following the methodology described above.  We then ask for the quantile of the true value of $\lambda$ within this posterior.  Figure \ref{fig:ppplot} shows the cumulative distribution of this quantile value, the so-called p-p plot.  If posteriors are self-consistent, we expect the truth to fall within the X\% Bayesian credible interval X\% of the time, i.e., the p-p plot should be diagonal \citep[e.g.,][]{Cook2006,Sidery:2013,Veitch:2014}.   We confirm that the p-p plot is consistent with the diagonal within statistical fluctuations.

Having tested the method and its implementation, we now analyse the uncertainty in the inferred value of $\lambda$.  This time, we fix the value of $\lambda$ at $\lambda=0.05$ when generating mock data catalogs, but vary the number of simulated events, with a subset of the events labeled as detectable.  We compute the width of the 90\% credible interval on $\lambda$, defined here as stretching from the 5th to the 95th percentile of the posterior.  In figure, we plot this width $\Delta \lambda$ against the number of detectable events.

\begin{figure}\center
  \includegraphics[width=1.05\columnwidth]{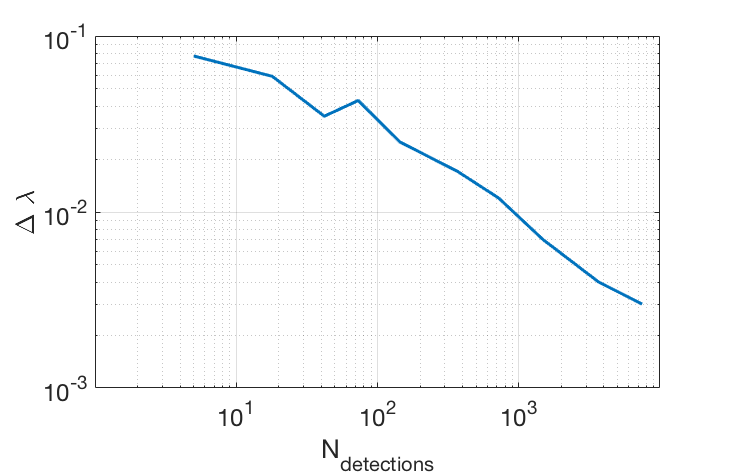}
  \caption{The width of the 90\% credible interval $\Delta \lambda$ as a function of the number of detections; the true value of $\lambda$ is 0.05 in all mock catalogs.  The fluctuations relative to the $\Delta \lambda \propto \Ndet^{-1/2}$ trend are due to the stochastic nature of the detected sample.}
\label{fig:deltalambda}
\end{figure}

We find that $\sim 1000$ detections at $\rho \geq 18$ are necessary in order to measure $\lambda$ to an accuracy $\delta \lambda \approx 0.01$.  Distributions with $\lambda=0.01$ and $\lambda=0.02$ yield median values of $\eta$ ($q$) of 0.243 (0.71) and 0.236 (0.62), respectively, so at least a thousand detections are required in order to make meaningful inference on the mass ratio distribution with a view to distinguishing evolutionary models.  An even greater number of detections would be required in each of several redshift bins in order to search for redshift-dependent changes in the mass ratio distribution -- perhaps $\mathcal{O}(10000)$, given the plausible variation of the mass ratio distribution with redshift.

\section*{Acknowledgments}
IM and WF thank Tom Loredo for useful discussions and the Statistical and Applied Mathematical Sciences Institute, partially supported by the National Science Foundation under Grant DMS-1127914, for hospitality. IM's work was performed in part at Aspen Center for Physics, which is supported by National Science Foundation grant PHY-1607611; IM's visit there was partially supported by a grant from the Simons Foundation.  IM thanks Stephen Justham, Vicky Kalogera, and Fred Rasio for discussions related to the illustrative example.  We thank Arya Farahi for alerting us to a typo in the manuscript, and the anonymous referee for a number of insightful comments.

\bibliographystyle{mnras}
\bibliography{Mandel}

\label{lastpage}
\end{document}